\documentclass{PoS}
\usepackage{amsmath}
\usepackage{color}
\definecolor{David}{rgb}{0.0,1.0,0.0}

\definecolor{Jason}{rgb}{1.0,0.0,0.0}

\title{Matrix elements from moments of correlation functions}

\ShortTitle{Moments of correlation functions}

\author{\speaker{Chia Cheng Chang}\textsuperscript{\textnormal{\textit{a}}}, 
Chris Bouchard\textsuperscript{\textnormal{\textit{b}}},
Kostas Orginos\textsuperscript{\textnormal{\textit{b},\textit{c}}},
David Richards\textsuperscript{\textnormal{\textit{c}}}\\
    \textsuperscript{a}Lawrence Berkeley National Laboratory\thanks{Lawrence Berkeley National Laboratory is operated by the DOE, Office of Science, Office of Nuclear Physics under Contract No. DE-AC02-05CH11231.}\\ 
    \textsuperscript{b}The College of William \& Mary\\
    \textsuperscript{c} Thomas Jefferson National Accelerator Facility\thanks{Authored by Jefferson Science Associates, LLC under U.S. DOE Contract No. DE-AC05-06OR23177. The U.S. Government retains a non-exclusive, paid-up, irrevocable, world-wide license to publish or reproduce this manuscript for U.S. Government purposes.}\\
    \\
        E-mail: \email{chiachang@lbl.gov}}

\abstract{Momentum-space derivatives of matrix elements can be related to their coordinate-space moments through the Fourier transform. We derive these expressions as a function of momentum transfer $Q^2$ for asymptotic in/out states consisting of a single hadron.  We calculate corrections to the finite volume moments by studying the spatial dependence of the lattice correlation functions.  This method permits the computation of not only the values of matrix elements at momenta accessible on the lattice, but also the momentum-space derivatives, providing {\it a priori} information about the $Q^2$ dependence of form factors. As a specific application we use the method, at a single lattice spacing and with unphysically heavy quarks, to directly obtain the slope of the isovector form factor at various $Q^2$, whence the isovector charge radius. The method has potential application in the calculation of any hadronic matrix element with momentum transfer, including those relevant to hadronic weak decays.}

\FullConference{34th annual International Symposium on Lattice Field Theory\\
		24-30 July 2016\\
		University of Southampton, UK}

\begin{document}

\section{Introduction}
Direct calculation of momentum-dependent slopes of hadronic form factors allows for model independent determinations of physical observables such as the charge radius of the proton.  Current experimental tension of the proton charge radius stand at a startling $7\sigma$ between electron and muonic measurement~\cite{Carlson:2015jba}. A model independent lattice QCD calculation of the charge radius at the 2\% level is necessary to discriminate the 4\% difference between electron and muon probes. Similarly, there is a $2\sigma$ tension between the nucleon axial mass, parameterising the $Q^2$ dependence of the nucleon axial form factor $F_A(Q^2)$ under a dipole ansatz and related to the inverse of the axial radius, obained from quasi-elastic scattering and electroproduction experiments~\cite{Anikin:2016teg}. The nucleon axial mass is the leading uncertainty when interpreting neutrino scattering experiments in the quasi-elastic regime. A direct lattice QCD calculation of $F_A(Q^2)$ with momentum transfer up to $Q^2=100~\text{GeV}^2$, in conjunction with the corresponding  derivatives with respect to $Q^2$  may be used together in a model independent $z$-expansion~\cite{Bhattacharya:2011ah} in order to constrain $F_A(q^2)$.  The uncertainty of 1\% for $F_A(0)$, required to interpret the next generation neutrino experiments such as the Deep Underground Neutrino Experiment (DUNE), is unobtainable with current lattice techniques and  available computational resources~\cite{Bhattacharya:2016zcn}. At higher levels of precision, isospin and electromagnetic corrections dominate. Similarly, in conjunction with precision flavor physics experiments, lattice QCD calculations of the shape of form factors for semi-leptonic decays such as $B_s\rightarrow K \ell \nu$~\cite{Bouchard:2014ypa}, are used to constrain Standard Model parameters such as $|V_{ub}|$. The dominant error at large momentum transfer comes from the uncertainty in the form factors. The additional information provided by explicit lattice calculations of the derivatives of form form factors may be used to reduce the uncertainties in the range of  $Q^2$ where form factors are computed.

Application of coordinate-space moment methods was first introduced in the mid 90's as a way to access the slope of the Isgur-Wise function $\xi(\omega)$ at zero-recoil, in order to interpret experimental results near zero-recoil from $B\rightarrow D$ semi-leptonic decays to be used to provide a value for $|V_{cb}|$~\cite{Lellouch:1994zu}. At the turn of the millenium, moment methods were used to calculate the slope of the energy-momentum tensor form factor, which is directly related to the the angular momentum contribution to the spin of the nucleon~\cite{Mathur:1999uf}\cite{Gadiyak:2001fe}. Recently, there is revitalized interest in coordinate-space moment methods both in applications to hadronic vacuum polarization calculations~\cite{Chakraborty:2016mwy}\cite{Blum:2016xpd} in particle physics, and direct calculations of the anomolous magnetic moment of the nucleon and nucleon radii in nuclear physics~\cite{Alexandrou:2016rbj}. In parallel, momentum-space derivative methods are also being explored to access similar nucleon structure calculations such as the anomolous magnetic moment and various nucleon radii~\cite{deDivitiis:2012vs}\cite{Tiburzi:2014yra}.  In these proceedings, we present a coordinate-space method that directly calculates the slope of single particle form factors with respect to 3- and 4-momentum squared at any lattice accessible momenta.

\section{Formalism}
\begin{figure}[h]
	\centering
		\includegraphics[width=0.65\textwidth]{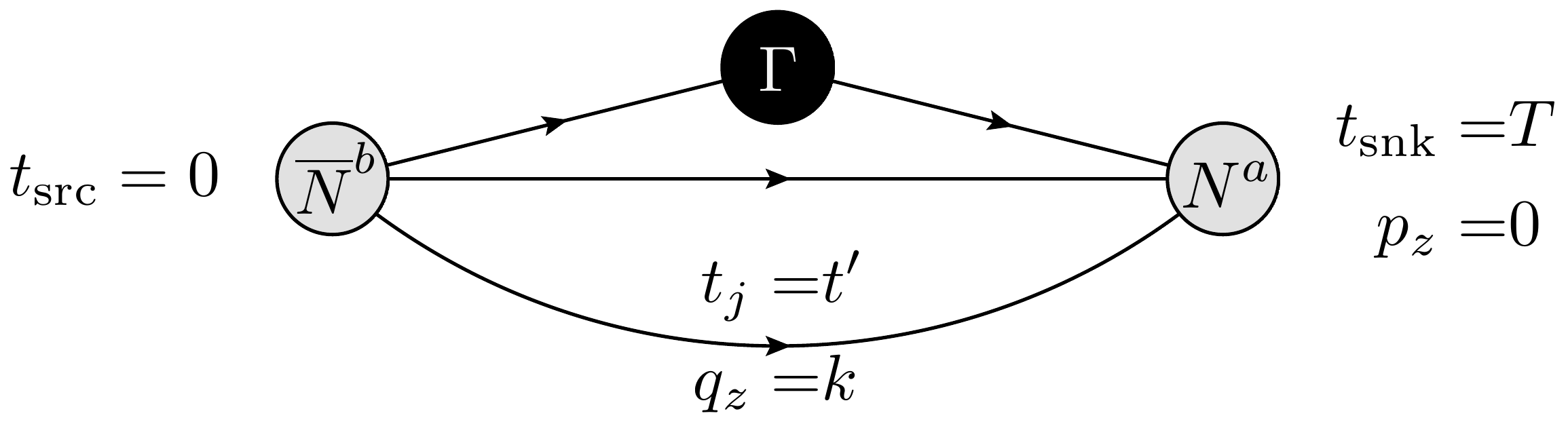}
	\caption{Kinematics of the three-point correlator with baryon initial and final states. We work in the rest frame of the final hadron. The diagram for semi-leptonic decays of mesons involves only one spectator quark, but involves the same kinematics.}
	\label{fig:3pt_kinematics}
\end{figure}

Given a three-point correlation function with the initial state at rest, and current insertion with three-momentum $k$, where $k$ points in the $z$-direction without loss of generality, as shown in Fig.~\ref{fig:3pt_kinematics}, the three-momentum-projected three-point has the general form,
\begin{align}
C^{\text{3pt}}(t, t^\prime) = \sum_{\vec{x},\vec{x}^\prime} \left<N^a_{t,\vec{x}}\Gamma_{t^\prime,\vec{x}^\prime} \overline{N}^b_{0,\vec{0}}\right> e^{-ikx^\prime_z},
\label{eq:3pt}
\end{align}
translational invariance allows us to shift the source to the origin $\vec{x}_{\text{src}} = 0$ and the sink to $\vec{x}\equiv \vec{x}_{\text{snk}} - \vec{x}_{\text{src}}$. Since the sink has zero three-momentum, the only momentum dependence left is at the current insertion. The operators $N^b$ and $N^a$ are the source and sink interpolation operators respectively, while the superscripts $a$ and $b$ label the nucleon interpolating operators~\cite{Basak:2005ir,Basak:2007kj}. The operator $\Gamma$ is a generic current insertion at position $\vec{x}^\prime\equiv \vec{x}_J-\vec{x}_{\text{src}}$.

The derivative of the three-point correlator with respect to $k^2$ follows,
\begin{align}
C^\prime_{\text{3pt}}(t, t^\prime)
%\equiv &\frac{\partial}{\partial k^2}C_{\text{3pt}}(t, t^\prime),\nonumber\\
=& \sum_{\vec{x},\vec{x}^\prime} \frac{-x^\prime_z}{2k}\sin\left(kx^\prime_z\right) \left<N^a_{t,\vec{x}}\Gamma_{t^\prime,\vec{x}^\prime} \overline{N}^b_{0,\vec{0}}\right>, 
\label{eq:3ptmoment}
\end{align}
where in Eq.~(\ref{eq:3ptmoment}), the cosine component vanishes due to symmetry. In the limit of zero momentum, the $k^2 \rightarrow 0$ limit of the integrand is given by L'H\^opital's rule,
\begin{align}
\lim_{k^2 \rightarrow 0} C^\prime_{\text{3pt}}(t, t^\prime) = & \sum_{\vec{x},\vec{x}^\prime}\frac{-x^{\prime 2}_z}{2}\left<N^a_{t,\vec{x}}\Gamma_{t^\prime,\vec{x}^\prime} \overline{N}^b_{0,\vec{0}}\right>.
\label{eq:3ptmoment0}
\end{align}

Given a two-point correlator with three-momentum $k$ in the $z$-direction,
\begin{align}
C_{\text{2pt}}(t) = \sum_{\vec{x}} \left< N^b_{t,\vec{x}}\overline{N}^b_{0,\vec{0}}\right> e^{-ikx_z},
\label{eq:2pt}
\end{align}
we can derive the derivative of the two-point correlator with respect to $k^2$,
\begin{align}
C^\prime_{\text{2pt}}(t)
% \equiv & \frac{\partial}{\partial k^2} C_{\text{2pt}}(t),  \nonumber\\
= & \sum_{\vec{x}}\frac{-x_z}{2k} \sin\left(kx_z\right) \left<N^b_{t, \vec{x}}\overline{N}^b_{0,\vec{0}} \right> ,
\label{eq:2ptmoment}
\end{align}
where analogous to the moment of the three-point correlator, the cosine contribution vanishes due to symmetry.  Consequently, in the zero-momentum limit,
\begin{align}
\lim_{k^2\rightarrow 0 } C^\prime_{\text{2pt}}(t) = \sum_{\vec{x}} \frac{-x_z^2}{2}\left<N^b_{t, \vec{x}}\overline{N}^b_{0,\vec{0}} \right>.
\label{eq:2ptmoment0}
\end{align}
The construction of the moment of the two-point correlator is similar to the moment of the three-point correlator with the exception that the moment now depends on the final state position $x_z$ instead of the current insertion position $x^\prime_z$. In both cases, the spatial moments at all momenta are even, resulting in a non-vanishing correlator under the Fourier transform, explicitly circumventing the concern raised in Ref~\cite{Wilcox:2002zt}.  Given prior computational investment in generating the propagators and sequential propagators, the generation of the moments of correlators only differ during the Fourier transform, and therefore require negligible additional computing time to construct.

\subsection{Interpretation}
In the rest frame of the final hadron, the time behaviour of the three-point correlator of Eq.~(\ref{eq:3pt}) is given by,
\begin{align}
C^{\text{3pt}}(t, t^\prime)  = & \sum_{n,m} \frac{Z_n^{\dagger a}(0) \Gamma_{nm}(k^2)Z_m^b(k^2)}{4 M_n(0) E_m(k^2)} e^{-M_n(0)(t - t^\prime)} e^{-E_m(k^2) t^\prime} \label{eq:3ptfit},
\end{align}
defining
\begin{align}
Z_n^{\dagger a}(0) &\equiv\left<\Omega | N^a | n, p_i=\left(0, 0, 0\right) \right>
\\
Z_m^b(k^2) &\equiv \left<m, p_i=\left(0,0,k\right)\right| \overline{N}^b\left|\Omega\right>\\
\Gamma_{nm}(k^2) &\equiv \left<n,p_i=\left(0,0,0\right)| \Gamma | m, p_i=\left(0,0,k\right) \right>
\end{align}
where $M_n(0)$ as the rest mass of the $n$-th state of the, and $E_m(k^2) = \sqrt{M_m^2+k^2}$ is the energy. The initial and final states created from and annihilated into the vacuum $\left|\Omega\right>$ have overlap with an infinite tower of states labelled by $n$ and $m$ with corresponding eigenvalues $M_n$ and $E_m$ respectively. We have performed a Wick rotation to imaginary time, yielding a sum of terms that decay exponentially in time. At large time separations, we recover the ground state signal.

Taking the derivative of Eq.~(\ref{eq:3ptfit}) with respect to $k^2$ yields,
\begin{align}
C^\prime_{\text{3pt}}(t, t^\prime) = &\sum_{n,m} C^{\text{3pt}}_{nm}(t,t^\prime)\left\{ \frac{\Gamma_{nm}^\prime(k^2)}{\Gamma_{nm}(k^2)} + \frac{Z^{b\prime}_m(k^2)}{Z^b_m(k^2)} - \frac{1}{2[E_m(k^2)]^2} -\frac{t^\prime}{2E_m(k^2)}\right\} \label{3ptmomentfit},
\end{align}
where $C_{nm}^{\text{3pt}}$ are the individual $n$ and $m$ contributions to the three-point correlator, and the prime denotes the first derivative with respect to three-momentum $k^2$. The derivative does not act on $Z^a$, which is in the rest frame of the initial particle regardless of momentum transfer at the current. The derivative only acts on $Z^b$ where the operator $N^b$ is constructed from non-local (smeared) interpolating operators, and thus has momentum dependence. We can obtain $Z_m^{b\prime}$ by looking at the moment of the two-point correlator constructed from the operator $N^b$, as we now show. We begin by using completeness to write the two-point correlator as
\begin{align}
C_{\text{2pt}}(t)= & \sum_m \frac{Z^{b\dagger}_m(k^2)Z^b_m(k^2)}{2E_m(k^2)}e^{-E_m(k^2)t} \label{eq:2ptfit},
\end{align}
where we project states to definite momentum $k$. Applying the $k^2$ derivative to the right-hand-side of Eq.~(\ref{eq:2ptfit}) yields
\begin{align}
C^\prime_{\text{2pt}}(t) 
%=\frac{\partial}{\partial k^2} \sum_m C_m^{\text{2pt}}(t)
= & \sum_m C^{\text{2pt}}_m(t) \left(\frac{2Z_m^{b\prime}(k^2)}{Z_m^b(k^2)} - \frac{1}{2[E_m(k^2)]^2}-\frac{t}{2E_m(k^2)}\right),
\end{align}
with respect to the four-momenta, $q^\mu\equiv (E_m(k) - M_n, 0, 0, k)$, such that $Q^2=-q^2$, where for a fixed three-momentum $k$ the four-momentum depends on the energies of the incoming and outgoing states. The connection between the three- and four-momentum derivatives arises from the chain rule,
\begin{align}
\frac{\partial}{\partial k^2} \Gamma_{nm} =  \frac{\partial Q^2}{\partial k^2}\frac{\partial}{\partial Q^2}\Gamma_{nm}
=  \frac{M_n}{\sqrt{(M_m)^2+k^2}}\frac{\partial}{\partial Q^2} \Gamma_{nm}.
\end{align}

\subsection{Remarks}
Correlators constructed from coordinate-space moment methods can have zero overlap with states at zero momentum, and in particular the zero-momentum ground state, and an example in  Ref.~\cite{Wilcox:2002zt} was instead shown to have overlap starting with the lowest non-zero lattice momentum mode. The subtlety lies in the fact that in Ref.~\cite{Wilcox:2002zt}, an odd spatial moment (or more specifically a linear moment) is taken in order to calculate the anomolous magnetic moment. Naively, as one takes the projection to zero momentum with an odd spatial moment, the correlator would vanish due to the oddness of the spatial integral.  However, the cancellation does not occur when an even spatial moment is constructed, and is in fact illustrated in a second example given in Ref.~\cite{Wilcox:2002zt}. We derive in the following sections the moment correlator and fit ansatz, and show that for all values of momenta, even spatial moments are constructed, yielding overlap with the desired ground state.

\section{Charge radius of the proton}
We demonstrate our method by calculating the charge radius of the proton on the 2+1 flavor William and Mary / JLab isotropic clover ensemble.  We use the 0.12~fm lattice spacing ensemble with a box size of $N_s^3\times N_t = 24^3\times 64$ with a pion mass of 400~MeV. To minimize the dominant error, which is the finite volume effects experience by the valence quarks, we work on a doubled lattice in the $z$-direction,  $N_{x,y}^2 \times N_z \times N_t = 24^2\times 48\times 64$.  A total of 480 configurations with 16 sources are analyzed in the follow results.

\subsection{Correlator analysis}
We perform a Bayesian constrained curve fit simultaneously to the two-point correlator, three-point correlators, and their respective moments. For the three-point correlator, we include two sink locations to disentangle excited state contributions. Furthermore, we work with Gaussian smeared source and sink operators to suppress these excited states. Finally, we only study the connected contribution to the vector form factor of the proton.

The ground state priors are motivated from looking at the standard effective mass and scaled correlator plots. The width of these ground state priors are orders of magnitude wider than the posterior distribution, rendering them effectively unconstrained. The excited state energies are defined as a tower of splittings with lognormal distributions where the mean is $2m_\pi$ and the width is of the same order of magnitude. Priors for the excited state overlap factors, matrix elements, and slopes are chosen to have a mean of zero, with a width that is five times the ground state central value.

\begin{figure}[h]
	\centering
		\includegraphics[width=0.8\textwidth]{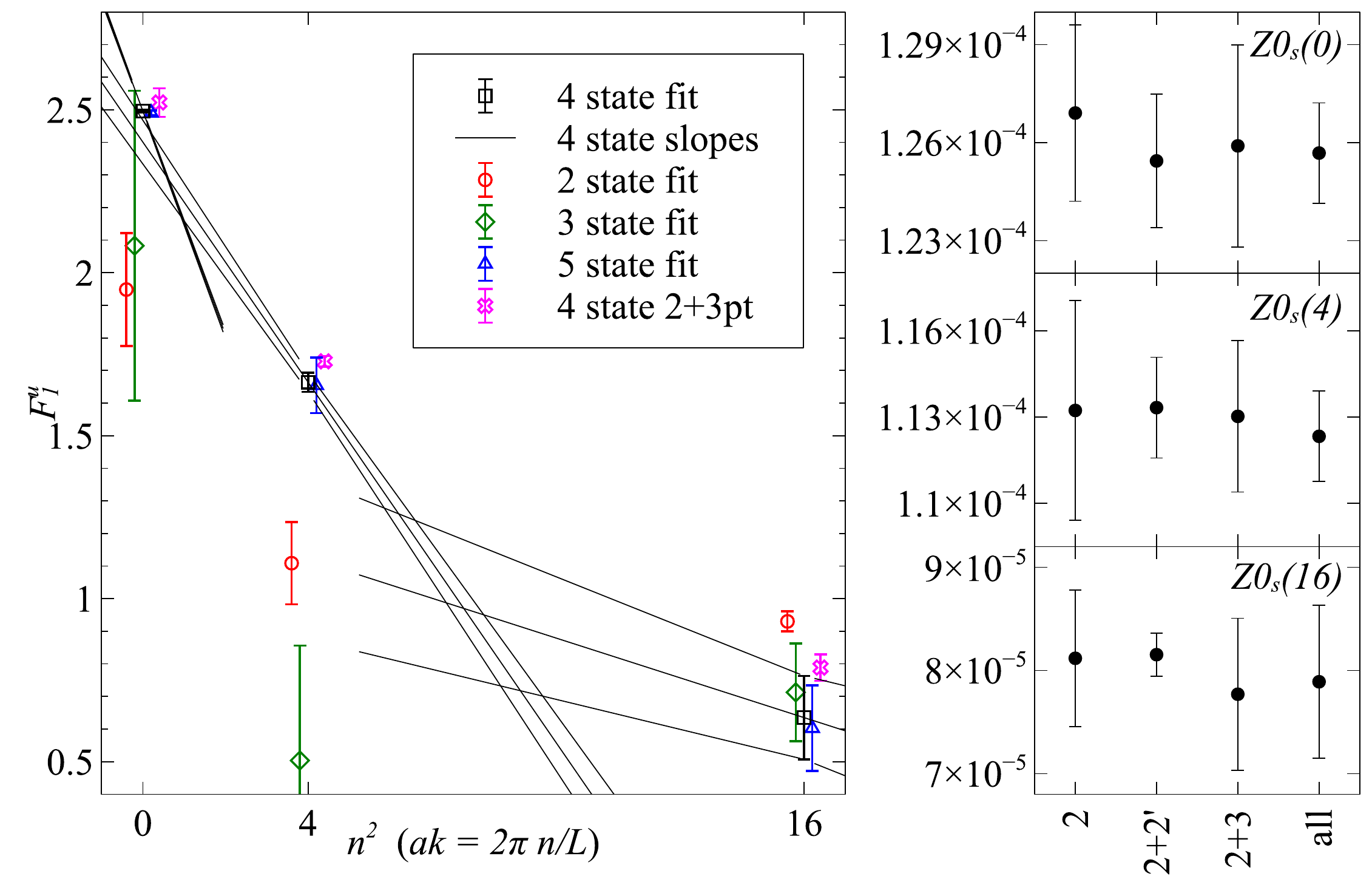}
	\caption{(Left) Fit of $F_1^u$, the connected contribution to the proton vector form factor. For all relevant fits, the two-point correlator fit region spans time separations of 2 to 10, the moment of the two-point correlator spans time separations of 3 to 11. The three-point correlator with $T_{\text{snk}}=10$ is fit with current insertion times of 2 to 8, while $T_{\text{snk}}=12$ has time separations spanning 2 to 10; the corresponding three-point moments fit regions span 4 to 8 and 4 to 9 for the two sink separations respectively. For the preferred fit with 4 states in the fit ansatz, the $\chi^2_\text{aug}/\text{d.o.f.}$ is 2.0, with the data contributing to 80\% of the total $\chi^2$; the augmented-$\chi^2$ is the sum of the $\chi^2$ contribution from data and the $\chi^2$ contribution from priors~\cite{Lepage:2001ym}. (Right) We check stability under fitting over subsets of data, for simultaneous fits of 2: exclusive two-point, $2+2^\prime$:two-point plus two-point moment, 2+3: two-point with three-point, all: two-point, two-point moment, three-point, and three-moment.}
	\label{fig:gV}
\end{figure}

We demonstrate full control over fit systematics by checking for stability under varying fit region, ground state prior widths and number of states included in the fit ansatz. In Fig.~\ref{fig:gV}, we plot our result for the vector form factor $F^u_1$ at three different lattice momenta, and show control of excited state systematics.  The preferred fit is marked by the black square, with corresponding slopes extracted from the simultaneous correlator fit. We vary the number of states in our fit ansatz from 2 (red circle), 3 (green diamond), and 5 (blue triangle), and observe that with 4 or more states, the central values of $F_1^u$ are stable. We further compare the value of $F_1^u$ against a simultaneous two- and three-point fit (pink cross), and observe consistency. The three panels on the right side of Fig.~\ref{fig:gV} compare the ground state overlap factors obtained from a two-point correlator only, two-point correlator with its moment, two- and three-point correlators, and finally the full simultaneous fit to all correlators.  The consistency of the ground state overlap factors between different subsets of data further demonstrate confidence in our fit analysis. We observe promising results from this method; the slopes for the three-values of lattice momenta agree well with expectation, and the uncertainty of the slopes are of the same order, but slightly larger, than the uncertainty on $F_1^u$.

\section{Outlook}
In these proceedings, we present a method to calculate slopes at arbitrary lattice momenta for matrix elements directly on the lattice.  Immediately interesting applications include calculating the isovector axial radius to obtain $F_A(Q^2)$ and the charge radius of the proton.  In particle physics, this method may help further constrain CKM phenomenology through stronger constraints on the shape of weak decay form factors. The proposed method is computationally cheap and  easy to implement, therefore is suitable for a wide range of physics applications.

\section{Acknowledgements}
The work was supported by the U.S. Department of Energy (DOE), Office of Science, Early Career Research Program under FWP Number NQCDAWL (CCC) and by the DOE, Office of Science, Office of Nuclear Physics under Contract No. DE-AC02-05CH11231 (CCC), DE-FG02-04ER41302 (CMB, KNO), and DE-AC05-06OR23177 (KNO, DGR).  Computations for this work were performed on the Cyclops Cluster at The College of William \& Mary, purchased with funds from the DOE, Office of Science, Early Career Research Program under FWP Number NQCDAWL, the resources of the National Energy Research Scientific Computing Center (NERSC), a DOE Office of Science User Facility supported by the Office of Science of the U.S. Department of Energy under Contract No. DE-AC02-05CH11231, and Extreme Science and Engineering Discovery Environment (XSEDE), which is supported by National Science Foundation grant number ACI-1053575.

\bibliographystyle{physrev}
\bibliography{bibliography}

\end{document}